\documentclass{article}
\usepackage{arxiv}
\usepackage{multirow}
\usepackage{booktabs}
\usepackage{cite}
\usepackage{amsmath,amssymb,amsfonts}
\usepackage{algorithmic}
\usepackage{graphicx}
\usepackage{textcomp}

\usepackage{verbatim} 
\usepackage{soul} 
\usepackage{wasysym} 

\usepackage[T1]{fontenc}
\usepackage[utf8]{inputenc}
\usepackage[emoticons]{emoji}
\usepackage{hyperref}

\title{A literature survey on student feedback assessment tools and their usage in sentiment analysis}

\author{
 Himali Aryal\\
 \textit{Department of Information Technology and Electrical Engineering} \\
    \textit{Norwegian University of Science and Technology}\\
    Gjøvik, Norway \\
    \href{mailto:himalia@stud.ntnu.no}{himalia@stud.ntnu.no} }

\begin{document}
\maketitle

\begin{abstract}

Online learning is becoming increasingly popular, whether for convenience, to accommodate work hours, or simply to have the freedom to study from anywhere. Especially, during the Covid-19 pandemic, it has become the only viable option for learning. The effectiveness of teaching various hard-core programming courses with a mix of theoretical content is determined by the student interaction and responses. In contrast to a digital lecture through Zoom or Teams, a lecturer may rapidly acquire such response from students' facial expressions, behavior, and attitude in a physical session, even if the listener is largely idle and non-interactive. However, student assessment in virtual learning is a challenging task. Despite the challenges, different technologies are progressively being integrated into teaching environments to boost student engagement and motivation. In this paper, we evaluate the effectiveness of various in-class feedback assessment methods such as Kahoot!, Mentimeter, Padlet, and polling to assist a lecturer in obtaining real-time feedback from students throughout a session and adapting the teaching style accordingly. Furthermore, some of the topics covered by student suggestions includes tutor suggestions, enhancing teaching style, course content, and other subjects. Any input gives the instructor valuable insight into how to improve the student's learning experience, however, manually going through all of the qualitative comments and extracting the ideas is tedious. Thus, in this paper we propose a sentiment analysis model for extracting the explicit suggestions from the students’ qualitative feedback comments.

\end{abstract}

    

    
    \section{Introduction}
\label{Introduction}

Online education has gained in popularity over the years due to various aspects such as accessibility in form of massive open online courses \cite{dalipi2016towards}, affordability and flexibility as result of distance and blended education \cite{imran2009interactive, garrison2009implications, kastrati2019spoc}, frameworks and tools \cite{imran2012multimedia, aparicio2016learning,  imran2016automatic,glancy2011conceptual}, interactivity  \cite{imran2014hip}, learning pedagogy and life-long learning \cite{Online_Education_attraction}. It has proven to be a most effective solutions for learning, especially during the COVID-19 pandemic. According to World Health Organization (WHO) data, COVID-19 has been recorded in over 216 countries, with millions of confirmed cases \cite{WHO}. Many countries have implemented precautionary measures, such as school and university lockdowns \cite{mukhtar2020advantages}. As a result, educational institutions began to provide the majority of their services online, including lectures and various examinations, to over 60\% of students throughout the world via various platforms \cite{elearning_in_lockdown, imran2014hip}. 
Online learning environments represent a distinct way of interaction. It entails the development of educational materials, delivery of instruction, and program management using the internet and other essential technologies \cite{what_is_online_learning}. During online learning, it is critical to maintain student interaction to curb dropouts \cite{dalipi2018mooc, marquez2016early, imran2019predicting}, and react to their responses while also obtaining feedback to/from them. Feedback and assessment are used to track student progress, manage learning pace, and assess instructional methods \cite{formative_assessment}. There are two kinds of student evaluations: summative and formative assessment. Summative assessment is a method of attempting to summarize student learning at a specific point in time, such as the end of a course. The majority of standardized examinations are in the form of a summative assessment. They aren't designed to give teachers and students quick, contextualized feedback during the learning process.  Formative assessment, on the other hand, occurs when teachers give students feedback in ways that help them learn more effectively, or when students can engage in a comparable, self-reflective process \cite{SA_FA_definition}. If the primary purpose of assessment is to support high-quality learning, then formative assessment ought to be understood as the most important assessment practice \cite{value_of_FA}.
 In a typical classroom, different characteristics such as students' faces and gestures assist the teacher in evaluating the behaviour \cite{pireva2015user} and sentiment of the students even though they are idle. However, formative assessment in virtual learning is a challenging task. Despite the challenges, different technologies are progressively being integrated into teaching environments to boost student engagement and motivation \cite{pireva2019evaluating, dalipi2017analysis}. Different student response systems have been utilized to improve student participation and real-time responses, including Mentimeter, Plickers, GoSoapBox, and Poll Everywhere \cite{{effect_of_menti_kahoot}}. In addition, in recent years, the usage of game-based applications such as Kahoot! in the classroom has grown in popularity \cite{popularity_of_game_based_learning}. Although the concept is not new, its application in the classroom in the context of play or competition has a direct impact on various aspects such as classroom participation and encouragement to the development of the activity, perception of the learning, commitment to the subject, and interest in deepening theoretical concepts \cite{gamebased_learning_benefits}. 
 
 Furthermore, student responses are also important for improving pedagogic and assessing students' learning behavior, and to provide personalized services and content via recommender systems \cite{imran2021systematic}. It helps to improve the interaction between teachers and students, and frequent input from students aids in the development of a better learning environment \cite{formative_assessment}. However, in most cases, feedback is provided after the course or a lecture in e-learning,  which might be too late to reconsider. Also, feedback can come from a different source, and it has to be analyzed to understand the opinion of students. Thus, in this literature based research paper we would focus on the effectiveness of different feedback assessment tools and how the responses from these tools can be utilized in the sentiment analysis to identify the student's attitude towards the course or teaching approach.

\subsection*{Research Questions}
    \begin{itemize}
        \item How many papers were published between 2015 to 2021 that have used digital tools to collect feedback and/or to perform quizzes? 
        \item What publications have these papers published in?
        \item Which tool has been explored more?
        \item When did the majority of the research take place?
        \item What are the most widely used evaluation metrics for feedback technologies that have been studied?
        \item How effective it is to adapt lectures and teaching style on the go during a lecture-based on students’ feedback?
        \item How sentiment analysis can be utilized in the responses collected from feedback assessment tools to identify the student's attitude towards the course or teaching approach?
    \end{itemize}


    \section{Related work and Literature survey}
\label{sec: Related Work}


There are currently a plethora of game-based learning tools available that use variations on game aspects to inspire, provide feedback, and structure participation pathways \cite{wang2016effect}. In this study, only the state-of-the-art of tools like Kahoot!, Mentimeter, and Padlet were explored. The bulk of research studies \cite{wang2019evaluation, raju2021effective, Padlet_in_Film_documentary, Game_Based_Digital_Quiz, gamebased_learning_benefits, wang2016effect, UcarKumte20178h, asee_peer_28276, novices2017Kahoot, licorish2018students} have looked at various aspects of these technologies, such as their effects on engagement, learning, classroom dynamics, concentration, motivation, and enjoyment, as well as how well they have been used in education. \cite{wang2020effect} Concluded that game-based learning has a positive impact on student performance. The students enjoy the lectures that include a quiz or survey in the middle. The studies \cite{wang2020effect, raju2021effective, Kahoot_effective_ACM, Mentimeter_intergration_model_CS_course, gamebased_learning_benefits, wang2016effect, asee_peer_28276, novices2017Kahoot} also demonstrated that there are numerous chances of losing concentration in e-learning, and these technologies aided to nurture concentration and increase engagement.

Additional focus of researches \cite{wang2020effect, mentimeter_in_elearning, bicen2018perceptions} were on how students and teachers perceived the use of these tools in education. The students' opinions of using these technologies in the classroom are also positive. According to the research \cite{mentimeter_in_elearning}, students noticed that these technologies helped them pay attention in class, and that Kahoot! or Mentimeter quizzes were particularly engaging. It has the ability to boost student attendance and has been regarded as interactive and quick. They also stated that the anonymity aspect allowed all students, including those who lacked confidence, to participate.

Furthermore, some studies \cite{wang2019evaluation, asee_peer_28276} compared the performance of students who utilized game-based learning to those who used followed traditional approaches. Their performance has greatly improved as a result of the findings. Students were divided into groups and taught using traditional methods such as PowerPoint presentations and encouraging students to ask questions and interact verbally, while others were taught using game-based learning, in which students played through a series of questions related to the topic while the teacher provided explanations in between. The evaluation \cite{Game_Based_Digital_Quiz} found that students who had the opportunity to learn in a game-based learning environment learned more than those who were taught using traditional methods. 

Another point of emphasis was the course teachers' real-time comments to the students. The \cite{mentimeter_in_elearning, Kahoot_effective_ACM, Mentimeter_intergration_model_CS_course, Padlet_in_Film_documentary, novices2017Kahoot, licorish2018students} studies looked at how the technologies studied could be used to interact with students and deliver immediate feedback based on their performance on a quiz or another activity. Only a few research, however, delved deeper into the analysis of feedback using machine learning techniques in order to use it to improve teaching style or other aspects of education.
In the research \cite{sentiment_on_online_poling} text pre-processing techniques were utilized to analyze student feedback obtained through various ways such as online polls and OMR sheets. According to the author, this experimental study of sentiment trees and feedback rating results in accurate polarities such as positive, negative, or neutral. Furthermore, the author stated that this technique would aid in determining better feedback findings, which faculty might use to improve their teaching process.

The table \ref{tab: effectiveness of feedback assessment tools} outlines the state-of-the-art of a few feedback assessment tools as well as summarizes the context in which these tools are assessed and aspects of which they have been studied.

    \begin{table}[!ht]
        \centering
        \caption{Literature survey of effectiveness of feedback assessment tools}
        \label{tab: effectiveness of feedback assessment tools}
        \begin{tabular}{p{1cm}p{2cm}p{1cm}p{2cm}p{6cm}p{4cm}}
         \hline
         R.N. & Author(s) & Year & Tool(s) & Context & Aspect \\ [0.8ex]
         \hline\hline 
          \cite{effect_of_menti_kahoot} &B. Gökbulut & 2020 & Kahoot!, Mentimeter  & To study effectiveness of using a gamification technique on e-learning & Effectiveness of tools in the e-learning\\
          
         \cite{wang2020effect} &A. I. Wang et al. & 2020 & Kahoot! & Study the advantages and disadvantages of using Kahoot in the classroom & Performance, and student and instructor attitudes \\
         
         \cite{mentimeter_in_elearning} & M. Mohin et al. & 2020 & Mentimeter & Working, its characteristics, applications and educational benefits, and students' perceptions of Mentimeter use & Formative assessment  \\
         
         \cite{bicen2018perceptions} &H. Bicen et al. & 2018 & Kahoot!  & Student attitude towards opinions gamification methods & Student motivation, effectiveness of using it classroom \\
         
         \cite{wang2019evaluation} &A. I. Wang et al. & 2019 & Kahoot!, PowerPoint, Sembly & Evaluated students' performance on three different teaching styles &  Interactiveness of gamified methodologies for teaching theoretical lectures \\
         
         \cite{UzunGaliKurb2020ke} &H.  Uzunboylu et al. & 2020 & Kahoot! & Evaluate the perception of teachers and students on the use of Kahoot! as a teaching tool  & Evaluated student's performance and semi-structured interview with teacher and student \\
         
         \cite{raju2021effective} &R. Raju et al. & 2021 & Kahoot!, Mentimeter, Quizizz &  Students engagement during online lectures & Student performance \\
        
         \cite{Kahoot_effective_ACM} &J. Catindig, M. S. Prudente & 2019 & Kahoot! & Effectiveness of Kahoot & Student performance, attitude \\
         
         \cite{Mentimeter_intergration_model_CS_course} &F. Quiroz Canlas et al. & 2019 & Mentimeter & Effectiveness of mentimeter in e-learning & Students' participation, engagement, interaction with instructor \\
         
          \cite{Cheatsheet_and_padlet} &F.  Soares et al. & 2020 & Cheat sheets, Padlet &  Effectiveness of tools &  Tool as a collaborative work \\
          
          \cite{Padlet_in_Film_documentary} &A.  R.  M.  Deni,  A.  Arifin & 2019 & Padlet & Evaluated Padlet as a repository which stores resources related to classroom activities and students' work-in-progress & Improved accessibility to teacher's input and peers' work-in-progress \\
          
          \cite{Game_Based_Digital_Quiz} & K. P. Nuci et al. & 2021 & Kahoot!, Google Form Quiz & In-lecture quizzes in online classes & Significant increase in students’ engagement, interaction, and interest in lectures \\
          
          \cite{gamebased_learning_benefits} & T.   Parra-Santos et al. & 2018 & Kahoot! & Quizzes, and Questionnaire & Students' attention, motivation and enjoyment are increased \\
          
          \cite{wang2016effect}  &A. I. Wang, A. Lieberoth & 2018 & Kahoot! & Use of points and audio affect in the learning environment  & Student engagement, learning, classroom dynamics, concentration, motivation and enjoyment \\
          
          \cite{UcarKumte20178h} &H.  Ucar,  A.  T.  Kumtepe & 2017  & Kahoot! & Qualitative case study is to explore the student's perspectives on game-based learning & Engage and motivate the online learners\\
          
          \cite{asee_peer_28276} & P . M. Tan, J. J. Saucerman& 2017 &  Kahoot!, SurveyMonkey &  Evaluated the benefits of gamification on students in the particular context of a student response system (SRS) & Student motivation, enjoyment, and encouragement to collaborate \\
          
          \cite{novices2017Kahoot} & C. M. Plump, J. LaRosa & 2017  &  Kahoot! & Quizzes & Student engagement and immediate feedback\\
          
          \cite{licorish2018students} & S. A. Licorish et al. & 2018 & Kahoot!  & Semi-structured interviews with students & Classroom dynamics, motivation and learning process, real-time feedback \\
          
           \hline
        \end{tabular}
    \end{table}

\subsection{Literature Review on Sentiment Analysis}
Sentiment analysis on student feedback has been an active research topic in the past years. The majority of the data utilized for the analysis comes from online courses, with only a few from the student feedback system. In this section we will highlight some of the research papers that uses student feedback in sentiment analysis. A survey table is also included at the end of the section, demonstrating the effectiveness of various feedback assessment tools.

The research work led by Z. Kastrati et al. \cite{Sentiment_Analysis_of_Students_Feedback_with_NLP_and_Deep_Learning} highlighted a systematic mapping of sentiment analysis on student feedback using NLP and Deep Learning. The research was conducted on 92 articles out of 612 that were found in the subject matter in learning platform environments. The PRISMA framework was utilized to guide the search procedure, which included only articles published between 2015 and 2020. The research identified that sentiment analysis on student's feedback using NLP and Deep Learning is a rapidly growing field, however, in order to fully mature the research and development in the field, structured datasets, standardized solutions, and work concentrated on emotional expression are essential. 

Another study \cite{deep_vs_traditional_on_feedback} presents a comparison of performances of 4 machine learning and natural language processing techniques: Naive Bayes, Maximum Entropy, Long Short-Term Memory, Bi-Directional Long Short-Term Memory for SA. According to the research, the Bi-Directional Long Short-Term Memory algorithm outperformed in terms of the F1-score measurement with 92.0\% on the sentiment classification task and 89.6\% on the topic classification task. In addition, a sentiment analysis application was developed that analyses student feedback and provides overview reports for the administrator to identify and recognize student interests in the institution.

The research study conducted in \cite{weakly_supervised_on_sentiment_analysis} investigated on aspect-level sentiment analysis to automatically identifies sentiment or opinion polarity of the student reviews. The weakly supervised tool was applied to effectively identify the aspect categories discussed in the unlabeled students' reviews. In addition, the results show that the framework performs admirably in terms of both aspect category identification and sentiment categorization. Furthermore, it is intended that the proposed framework would replace labor-intensive sentiment analysis techniques that heavily rely on manually labeled data. A similar research work is also conducted in \cite{kastrati2020aspect} where various traditional machine learning algorithms and deep neural networks are trained on real-world dataset composed of more than 21 thousands students' reviews manually classified into 5 different aspect categories and 3 polarity classes. Further improvements were made by applying deep learning models in \cite{edalati2021potential}. These models, however, relies only on the basic NLP techniques to process data for finding sentiments. To better capture domain specific sentiments for educational feedback, there is a need to incorporate both semantic and contextual information form users input. Such a model could utilize domain-specific ontology \cite{7050779,kastrati2016semcon, Choi200domain, kastrati2015improved}, and publicly available lexical resources such as SenticNet, SentiWordNet, etc., \cite{esuli-sebastiani-2006-sentiwordnet}. 

In the research \cite{sentiment_on_online_poling} text pre-processing techniques were utilized to analyze student feedback obtained through various ways such as online polls and OMR sheets. According to the author, this experimental study of sentiment trees and feedback rating results in accurate polarities such as positive, negative, or neutral. Furthermore, the author stated that this technique would aid in determining better feedback findings, which faculty might use to improve their teaching process.

    \section{Research Design}
\label{sec: Research Design}

\subsection{Search Strategy}
We used systematic mapping as a research methodology to review the literature for this research. To conduct this literature review, we followed the Preferred Reporting Items for Systematic Reviews and Meta-Analyses (PRISMA) statement \cite{PRISMA_STATEMENT}. The PRISMA method of systematic review enables a comprehensive review of all relevant articles and aids in the discovery of a clear response to the research question. The approach uses a variety of inclusion and exclusion criteria to help with decision-making.

  \begin{figure}[!ht]
      \centering
      \includegraphics[width=0.80\textwidth]{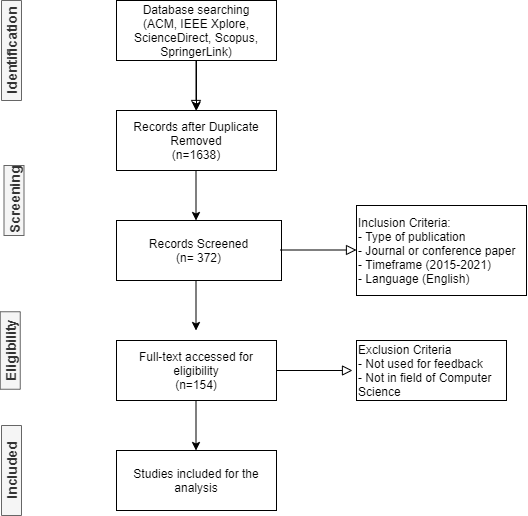}
      \caption{PRISMA Methodology}
      \label{fig: PRISMA Methodology}
    \end{figure}

For this study, we followed a desk work methodology where scientific papers that contained information feedback assessment tools were searched and gathered. The keywords search approach was used to locate these resources. Keywords like \lq Kahoot!', \lq Mentimeter', \lq Padlet',  \lq Feedback Assessment Tools, \lq Game-Based approach' were used to search these resources.

Following the initial research, we limited our search only to a few feedback assessment tools (\lq Kahoot!, \lq Mentimeter, \lq Padlet') and focused our survey on which context of these has been studied and which aspect i.e. information about the themes/topics for which students have commented on these tools are in the learning environment and whether they are beneficial to employ in the future. The table \ref{Query String used for searching articles} includes the query string used for searching the articles.
   \begin{table}[!ht]
   \centering
   \caption{Query String used for searching articles}
   \label{Query String used for searching articles}
    \begin{tabular}{l}
    \hline
    \textbf{Search   String (Query)}                                                                                                        \\
    \hline
    Kahoot!,      "Kahoot!" AND Feedback,      "Kahoot!" AND "Feedback" AND "Sentiment"          \\
     \hline
    Mentimeter,     "Mentimeter" AND "Feedback",      "Mentimeter" AND "Feedback" AND "Sentiment" \\
     \hline
    Padlet,      "Padlet" AND "Feedback",      "Padlet" AND "Feedback" AND "Sentiment"\\  
    \hline
    \end{tabular}
    \end{table}

\subsection{Time Period and Digital Databases}
The study incorporates the articles published between 2015 and 2021. Since the study was conducted in 2021, it encompassed papers published up until July of that year.
We used the following online research databases and engines to conduct our search:
\begin{itemize}
    \item ACM Digital Library
    \item IEEE Xplore
    \item ScienceDirect
    \item Scopus
    \item SpringerLink
\end{itemize}

\subsection{Identification of Primary Studies}
\begin{figure}[!ht]
      \centering
      \includegraphics[width=0.80\textwidth]{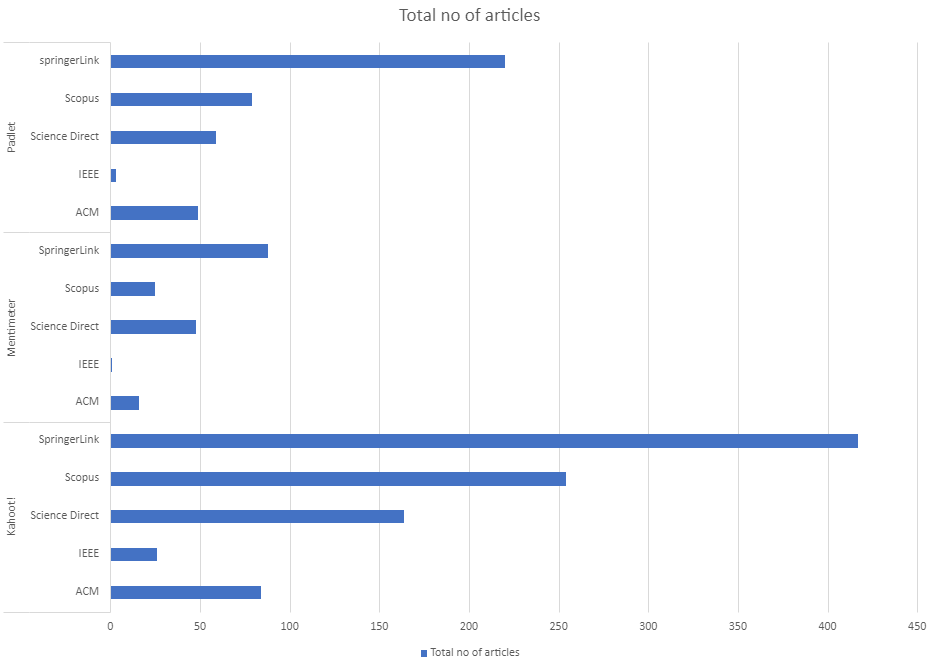}
      \caption{Number of articles found in different digital libraries}
      \label{fig: Number of articles found in different digital libraries}
    \end{figure}

The search began with a simple query, which we narrowed down sequentially. In the first stage, using the queries string \lq Kahoot!', \lq Mentimeter', and \lq Padlet', a total of 1638 articles published in the English language were gathered. The figure \ref{fig: Number of articles found in different digital libraries} depicts the statistics of all the articles acquired initially in the various digital libraries.

\subsection{Study Selection/Screening}
 The papers were screened using a number of inclusive criteria. Articles published only in English, papers released between 2015 and 2021, and prioritized peer-reviewed articles were among the inclusion criteria utilized for the filtering. The table \ref{tab:Articles published on different years} displays annual data for articles published on various feedback tools on different database sources.
 
 \begin{table}[!ht]
    \centering
    \caption{Articles published on different years}
    \label{tab:Articles published on different years}
    \begin{tabular}{|l|llllll|}
    \hline
    \textbf{Tools}              & \textbf{Year} & \textbf{ACM DL} & \textbf{IEEE Xplore} & \textbf{Science Direct} & \textbf{Scopus} & \textbf{SpringerLink} \\
    \hline
    \multirow{7}{*}{Kahoot}     & 2015          & 0               & 0                    & 4                       & 4               & 6                     \\
                                & 2016          & 3               & 1                    & 7                       & 7               & 11                    \\
                                & 2017          & 2               & 3                    & 13                      & 17              & 14                    \\
                                & 2018          & 20              & 9                    & 23                      & 49              & 60                    \\
                                & 2019          & 20              & 2                    & 24                      & 63              & 71                    \\
                                & 2020          & 31              & 5                    & 42                      & 73              & 99                    \\
                                & 2021          & 8               & 5                    & 51                      & 45              & 156                   \\
                                \hline
    \multirow{7}{*}{Mentimeter} & 2015          & 0               & 0                    & 0                       & 0               & 1                     \\
                                & 2016          & 0               & 0                    & 2                       & 0               & 1                     \\
                                & 2017          & 0               & 1                    & 1                       & 1               & 4                     \\
                                & 2018          & 3               & 0                    & 3                       & 3               & 4                     \\
                                & 2019          & 2               & 0                    & 7                       & 3               & 16                    \\
                                & 2020          & 6               & 0                    & 12                      & 12              & 19                    \\
                                & 2021          & 5               & 0                    & 23                      & 6               & 43                    \\
    \hline
    \multirow{7}{*}{Padlet}     & 2015          & 0               & 1                    & 3                       & 5               & 12                    \\
                                & 2016          & 2               & 1                    & 2                       & 4               & 9                     \\
                                & 2017          & 5               & 0                    & 9                       & 6               & 18                    \\
                                & 2018          & 5               & 0                    & 7                       & 12              & 31                    \\
                                & 2019          & 6               & 1                    & 9                       & 15              & 42                    \\
                                & 2020          & 20              & 0                    & 8                       & 26              & 51                    \\
                                & 2021          & 11              & 0                    & 21                      & 11              & 57                   \\
                                \hline
    \end{tabular}
\end{table}

\subsection{Eligibility Criteria}
 The search is narrowed down to specific sorts of articles, such as journal articles and conference papers. The research was also restricted to the subject of computer science. A total of 154 articles were identified to meet the eligibility requirements using the following inclusion criteria.
 \begin{itemize}
     \item Type of publication
     \item Field of study
     \item Journal or conference papers
     \item Time-frame (2015-2021)
     \item Language (English)
 \end{itemize} 

The table \ref{tab: Number of articles collected from various databases} shows the final 154 articles obtained with the above mentioned eligibility criteria. On the query string "sentiment AND feedback AND (mentimeter OR Kahoot! OR Padlet)" a few articles were retrieved; however, upon closer inspection, it was discovered that sentiment analysis was not done on the feedback received from the tools (Kahoot!, Mentimeter, and Padlet).
    
\begin{table}[!ht]
    \centering
    \caption{Number of articles collected from various databases}
    \label{tab: Number of articles collected from various databases}
    \begin{tabular}{|c|ccccc|}
    \hline
    Search   Queries & Digital Library            & No inclusion criteria & with inclusion & CS related & Journal/Conference \\
    \hline
                                                                 & ACM           & 68                         & 67             & 2          & 2                    \\
                                                                 & IEEE          & 7                          & 7              & 6          & 6                    \\
                                                                 & ScienceDirect & 116                        & 116            & 23         & 22                   \\
                                                                 & Scopus        & 40                         & 40             & 26         & 17                   \\
    \multirow{-5}{*}{Kahoot! AND Feedback}                       & SpringerLink  & 382                        & 346            & 106        & 56                   \\
    \hline                                                             
                                                                 & ACM           & 12                         & 12             & 1          & 1                    \\
                                                                 & IEEE          & 0                          & 0              & 0          & 0                    \\
                                                                 & ScienceDirect & 30                         & 30             & 3          & 2                    \\
                                                                 & Scopus        & 4                          & 4              & 3          & 2                    \\
    \multirow{-5}{*}{Mentimeter AND feedback}                & SpringerLink  & 72                         & 70             & 21         & 15                   \\
     \hline                                                             & ACM           & 0                          & 0              & 0          & 0                    \\
                                                                 & IEEE          & 1                          & 1              & 1          & 1                    \\
                                                                 & ScienceDirect & 40                         & 39             & 3          & 3                    \\
                                                                 & Scopus        & 15                         & 15             & 9          & 6                    \\
    \multirow{-5}{*}{Padlet AND feedback}                    & SpringerLink  & 212                        & 177            & 38         & 19  \\
    \hline 
    \textbf{Total} & & & & & \textbf{154}\\
    \hline
    \end{tabular}
\end{table}

    \section{Result and Discussion}
\label{Result and Discussion}
The answers to the research questions are included in this section.

\textit{How many papers were published between 2015 to 2021 that have used digital tools to collect feedback and/or to perform quizzes?}\\
Since the development of Kahoot!, it has become a popular research topic in the field of education, with a large number of articles published since then. In the Mentimeter and Padlet, it's the same. Without taking into account varied inclusion and exclusion criteria, we found a total of 1638 articles in these three tools. Whereas,  considering different inclusion/exclusion criteria, 154 articles were found in journals and conferences in the field of Computer Science between 2015 to 2021. These 154 articles were published either on Kahoot! or Mentimeter or Padlet in the context of feedback assessment or feedback analysis.

\textit{What publications have these papers published in?}\\
We focused our research on the ACM (Association for Computing Machinery), IEEE Xplore, ScienceDirect, Scopus, and SpringerLink digital libraries. SpringerLink is the most popular publisher (90 articles that meet the inclusion/eligibility criterion), while IEEE Xplore has the most list articles (7 items that meet the inclusion/eligibility criteria) as shown in the table \ref{tab: Number of articles collected from various databases}. Scopus is in third place, with 25 conference papers, after ScienceDirect, which has 27 conference/journal papers. 

\textit{When did the majority of the research take place?}\\
The total number of publications published in the year 2021 was the highest, meeting the inclusion criteria and eligibility criteria as shown in the table \ref{tab: Articles published per year}. After 2018, the number of articles published each year is higher than before. This increased number of articles in the recent years could be due to the COVID-19 outbreak. Because of the Covid-19 pandemic, educational institutions have been obliged to shift to an online teaching method all around the world \cite{mukhtar2020advantages}. As a result, an increased study on the topic of e-learning and various online feedback assessment systems was observed.

\begin{table}[!ht]
    \centering
    \caption{Articles published per year}
    \label{tab: Articles published per year}
    \begin{tabular}{|l|lllllll|}
    \hline
    Digital Library & 2015 & 2016 & 2017 & 2018 & 2019 & 2020 & 2021 \\
    \hline
    ACM             & 0    & 0    & 0    & 1    & 0    & 1    & 1    \\
    IEEE            & 0    & 0    & 1    & 1    & 1    & 0    & 1    \\
    ScienceDirect   & 2    & 2    & 1    & 3    & 6    & 5    & 10   \\
    Scopus          & 2    & 1    & 2    & 9    & 8    & 1    & 1    \\
    SpringerLink    & 2    & 4    & 7    & 19   & 21   & 17   & 23   \\
    \hline
    Total           & 6    & 7    & 11   & 34   & 36   & 25   & 37  \\
    \hline
    \end{tabular}
\end{table}

\textit{Which tool has been explored more?}\\
We looked into Kahoot!, Mentimeter, and Padlet, which are three different feedback assessment tools. According to the data, Kahoot! is the most extensively used tool. We discovered 103 articles solely on the Kahoot! tools with considering the inclusion/eligibility criteria. Whereas 29 studies focused on Padlet feedback analysis and 20 publications on mentimeter were identified. According to our observations, Kahoot! is the most popular tool among the technologies considered for feedback assessment in the learning environment.

\textit{What are the most widely used evaluation metrics for feedback technologies that have been studied?}\\
The 154 articles for the metadata analysis were chosen once the final inclusion and exclusion criteria were determined. Upon closer inspection, we discovered that studies had been undertaken in the following areas:
\begin{itemize}
    \item Student perception  
    \item Collaborative knowledge building
    \item Improve student learning motivation and performance
    \item Effectiveness of tool(s)
\end{itemize}

   

\textit{How effective it is to adapt lectures and teaching style on the go during a lecture-based on students’ feedback?}
The implementation of a feedback system in the lecture is seen positively by both students and teachers \cite{bicen2018perceptions,effect_of_menti_kahoot, mentimeter_in_elearning, UzunGaliKurb2020ke}. The perspectives and experiences of others who have used these tools in the classroom offer excellent insight all the while diversifying the learning process \cite{UzunGaliKurb2020ke}.

    \section{Identified Challenges and Research Gap}
\label{Identified challenges and research gap}

According to the results of the survey, several in-class feedback assessment tools such as Kahoot!, Mentimeter, Padlet are quite effective. However, we were unable to locate any research that focused on sentiment analysis on data collected from feedback tools. We can't deny that sentiment analysis research on student review isn't new; nonetheless, the most of studies have focused on data acquired from Coursera reviews \cite{weakly_supervised_on_sentiment_analysis}, social media pages \cite{sentiment_on_facebook_review}, or other online courses platforms \cite{kastrati2020wet}. The research on the sentiment analysis based on the feedback assessment tools especially Kahoot!, Mentimeter and Padlet is mostly unexplored.

\subsection{Future Direction} 
\label{Future Direction}

One approach to determine the polarity of student comments received via feedback assessment tools (Kahoot!, Mentimeter, and Padlet) is using sentiment analysis \cite{edalati2021potential}.
The methodology of Student Feedback Sentiment Analysis is depicted in the flowchart as shown in the figure \ref{fig: Proposed Sentiment Analysis Model}. The depicted model is inspired from \cite{imran2020cross,imran2020cross1}. We will focus on the following aspects when conducting sentiment analysis:

    \begin{figure}[!ht]
      \centering
      \includegraphics[width=0.80\textwidth]{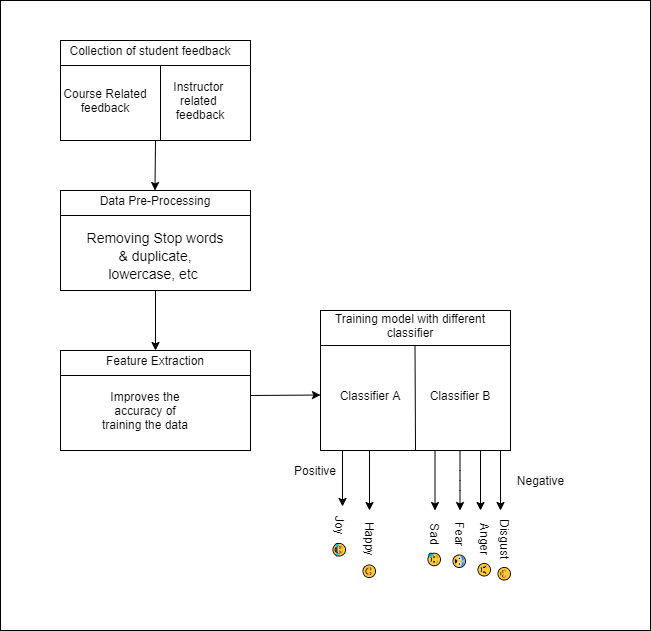}
      \caption{Proposed Sentiment Analysis Model}
      \label{fig: Proposed Sentiment Analysis Model}
    \end{figure}

\begin{enumerate}
    \item \textbf{Data Collection:}
    Data collecting is an important phase in the Sentiment Analysis. The more data there is, the better the result. In a course during the Fall semester of 2021, testing will be conducted for a short-list of tools such as Kahoot!, Mentimeter, and Padlet.
    
    \item \textbf{Data pre-processing:}
    Data consist of the mixture of text, numbers, and special characters. Data pre-processing is required before further data processing, including converting text to lowercase, reducing stop words, removing unnecessary punctuation, locating exclamation marks or question marks, and removing inconsistent letter casing.
    
    \item \textbf{Feature Extraction and Conceptualization of Sentiment:}
    We can better interpret the text by employing feature extraction, which will improve the accuracy of training the data using a new method. Term Presence and Frequency, Part of Speech Tagging, and Negation are some of the features that can be used. Also incorporating the semantic context using publicly available lexical databases (i.e WordNet, SentiWordNet, SenticNet, etc.) \cite{esuli-sebastiani-2006-sentiwordnet} or semantically rich representations using ontologies \cite{kastrati2019impact,THAKOR2015199} and their thesaurus \cite{kastrati2019performance,Priya2020multilevel} to identify opinion and attitude of users from text would be an import aspect to further investigate.
    
    \item \textbf{Training Model with different classifiers:}
    Following feature extraction, we can use a number of existing algorithms to model our data, including Naive Bayes, Support Vector Machines, CNN, and others.  The algorithm depends on the features of the data and there is no said algorithm that is perfect for any data. To identify the optimal model for the analysis, we'll run the data through a variety of features and algorithms.
    
\end{enumerate}

    \section{Conclusion}
\label{sec: Conclusion}
Due to the increasing demand of online education and distant learning as a result of the COVID19 epidemic, students’ feedback analysis is the most vital task for professors as well as educational institutes. Hence implementing various feedback assessment tool is essentials. Although a lot of study has been done on the effectiveness of feedback assessment tools, there is still work to be done on sentiment analysis on data received from feedback assessment tools (Kahoot!, Mentimeter, and Padlet). This article focused on the current state of research on feedback assessment tools, as well as the contexts in which these tools have been investigated and the status of sentiment analysis. The findings imply that feedback evaluation tools are now vital in the field of education, and that sentiment analysis on data acquired from these feedback assessment tools is rarely investigated.


    
    \bibliography{access.bib}{}
    \bibliographystyle{IEEEtran}

\end{document}